\documentclass[prl,aps,twocolumn]{revtex4}
\usepackage{epsfig,latexsym}

\begin{document}

\title{Probability distribution of the maximum of a smooth temporal signal}

\author{Cl\'ement Sire}

\affiliation{Laboratoire de Physique Th\'eorique (UMR 5152 du CNRS),
Universit\'e Paul Sabatier, 118, route de Narbonne, 31062 Toulouse
Cedex 4, France
\\
E-mail: {\it clement.sire@irsamc.ups-tlse.fr}}

\begin{abstract}
We present an approximate calculation for the distribution of the
maximum of a smooth stationary temporal signal $X(t)$.  As an
application, we compute the persistence exponent associated to the
probability that the process remains below a non-zero level $M$.
When $X(t)$ is a Gaussian process, our results are expressed
explicitly in terms of the two-time correlation function,
$f(t)=\langle X(0)X(t)\rangle$.
\end{abstract}

\maketitle

The problem of evaluating the distribution of the maximum of a
time-correlated random variable $X(t)$ has elicited a large body of
work by mathematicians \cite{adler,BL,merca}, and physicists, both
theorists \cite{SM,AB1,BD1,per1,per2,per3,iia,global,krug} and
experimentalists \cite{breath,lq,soap,diff1d,surf}. In the physics
literature, this is related to the persistence problem, the
probability that a temporal signal $X$ (and hence its maximum)
remains below a given level $M$ up to time $t$. Persistence
properties have been measured in as different systems as breath
figures \cite{breath}, liquid crystals \cite{lq}, laser-polarized
$Xe$ gas \cite{diff1d}, fluctuating steps on a $Si$ surface
\cite{surf}, or soap bubbles \cite{soap}.

The mathematical literature has mainly focused on evaluating
$P_<(t)={\rm Prob}(X(t')<M,\quad t'\in[0,t])$ for Gaussian processes
and for large $|M|$, a regime where efficient bounds or equivalent
have been obtained \cite{adler,BL}. Recently \cite{merca}, and for
Gaussian processes only, a numerical method to obtain valuable
bounds has been extended to all values of $M$, although the required
numerical effort can become quite considerable for large $t$.

Physicists have also concentrated their attention to Gaussian
processes \cite{per1,per2,per3,iia}, which are often a good or {\it
exact} description of actual physical processes. For instance, the
total magnetization in a spin system \cite{global}, or the height
profile of certain fluctuating interfaces \cite{krug,diff1d,surf}
are true temporal Gaussian processes. Two general methods have been
developed, focusing on the case $M=0$, which applies to many
physical situations. The first one \cite{per1,per2,per3} is a
perturbation of the considered process around the Markovian Gaussian
process, which has been extended for small values of $M$
\cite{per2}. Within this method, only the large time asymptotics of
$P_<(t)$ is known, leading to the definition of the persistence
exponent (see below). The alternative method, using the independent
interval approximation \cite{iia}, gives very accurate results for
smooth processes, but is restricted to $M=0$.

In addition, this problem has obvious applications in many other
applied and experimental sciences, where one has to deal with data
analysis of complex statistical signals. For instance, statistical
bounds of noisy signals are extremely useful for image processing
(for instance in medical imaging or astrophysics \cite{image}), in
order to obtain cleaner images by correcting spurious bright or dark
pixels \cite{adler,merca}. In general, it is important to be able to
evaluate the maximum of a correlated temporal or spatial signal
originating from experimental noise. The same question can arise
when the signal lives in a more abstract space. For instance, in the
context of genetic cartography, statistical methods to evaluate the
maximum of a complex signal has been exploited to identify putative
quantitative trait loci \cite{lander}. Finally, this same problem
arises in econophysics or finance, where the probability for a
generally strongly correlated financial signal to remain below or
above a certain level is of great concern.

One considers a {\it general} ({\it i.e.} not necessarily Gaussian)
stationary process $X(t)$ of distribution $g(X)$, and zero mean. The
process is assumed to be ``smooth'', so that its velocity $X'(t)$ is
continuous, ensuring that the number of times $N(t)$, where $X=M$,
remains finite for any bounded time interval $[0,t]$. For a given
level $M$, one defines $\tau$ as the average temporal interval
between two crossings of the level $X=M$. One also introduces
$P_-(t)$ (respectively $P_+(t)$) as the distribution of time
intervals during which $X(t)\leq M$ (resp. $X(t)\geq M$). The
average of $P_\pm(t)$ are denoted by $\tau_\pm$. Finally, one
defines $P_<(t)$ (resp. $P_>(t)$) as the probability, starting at
$X(0)<M$ (resp. $X(0)>M$), that the process $X$ remains below the
level $M$ (resp. above the level $M$) up to time $t$. The
distribution of the maximum (resp. minimum) of the process $X(t)$ in
the interval $[0,t]$ is clearly the derivative of $P_<(t)$ (resp.
$P_>(t)$) with respect to $M$. The difficulty of obtaining analytic
forms for the above quantities lies in the fact that powerful
methods like the Fokker-Planck approach are useless for
non-Markovian processes. In this letter, we obtain closed
expressions for $P_<(t)$, $P_>(t)$, and $P_\pm(t)$ from a minimal
knowledge of the statistical properties of the process $X$. This is
achieved by analyzing the trajectories of $X$ and using the sole
assumption that the lengths of the intervals between successive
crossings of the level $M$ are uncorrelated.

Let us assume that one {\it knows} the two following quantities
$A(t)$ and $N_<(t)$ from experiment, numerical simulations, or even
analytically: $A(t)$ is the autocorrelation function of $\theta
[M-X(t)]$ ($\theta$ is Heaviside's function),
\begin{equation}
A(t)=\langle \theta [M-X(t)]\, \theta [M-X(0)]\rangle,
\end{equation}
and $N_<(t)$ is the average number of crossings at level $M$ up to
time $t$, averaged over the starting position $X(0)<M$. For large
time, one has
\begin{equation}
N_<(t)\sim N(t)=\frac{t}{\tau},\label{as}
\end{equation}
with
\begin{equation}
N(t)=\left\langle
\int_0^t|X'(t')|\delta(X(t')-M)\,dt'\right\rangle.
\end{equation}
If the process is smooth, $X'(t')$ is not correlated with $X(t')$,
and using stationarity, we find $N(t)=t{\times}\langle
|X'(t)|\rangle\langle \delta(X(t)-M)\rangle$, which leads to
\begin{equation}
\tau^{-1}=g(M)\langle |X'(t)|\rangle.\label{taugen}
\end{equation}
In addition, $N_>(t)$, the average number of crossings at level $M$
up to time $t$, starting from $X(0)>M$, satisfies the sum rule
\begin{equation}
G(M)N_<(t)+(1-G(M))N_>(t)=\frac{t}{\tau},\label{sumrule}
\end{equation}
where $G(M)=\int_{-\infty}^M g(x)\,dx$. Note that $\tau_\pm$ are
simply related to $\tau=\frac{\tau_++\tau_-}{2}$:
\begin{eqnarray}
\tau_-=2\tau G(M),\quad\tau_+=2\tau (1-G(M)).\label{t2}
\end{eqnarray}

In the following, we obtain closed forms for $P_<(t)$, $P_>(t)$, and
$P_\pm(t)$ from the knowledge of $A(t)$ and $N_<(t)$, for any level
$M$. When $X$ is a Gaussian process of correlator $f(t)$, we shall
see later that $A(t)$, $N_<(t)$, $N_>(t)$, $\tau$, and $\tau_\pm$
can be explicitly written in terms of $f$. Hence, the minimal
knowledge of the two-time correlation function of a Gaussian process
will grant access to the yet unknown quantities $P_<(t)$, $P_>(t)$,
and $P_\pm(t)$. However, the present approach has a wider range of
applications and {\it does not rely} on the Gaussian property of the
process.

Our central approximation consists in assuming that the {\it
interval length between crossings are uncorrelated} \cite{iia}. The
probability $P_<(N,t)$, starting from $X(0)<M$, that there are
exactly $N$ crossings in the interval $[0,t]$, can then be written,
for odd $N=2n-1$ $(n\geq 1)$,
\begin{eqnarray}
&P_<(2n-1,t)={\tau_-^{-1}}\int_0^tdt_1\,Q_-(t_1){\times}
\nonumber\\
&\int_{t_1}^t dt_2\,P_+(t_2-t_1) \int_{t_2}^t
dt_3\,P_-(t_3-t_2)\cdots \nonumber\\
&\int_{t_{2n-3}}^t dt_{2n-2}\,P_+(t_{2n-2}-t_{2n-3}){\times}
\nonumber\\
&\int_{t_{2n-2}}^t
dt_{2n-1}\,P_-(t_{2n-1}-t_{2n-2})Q_+(t-t_{2n-1}),\label{conv1}
\end{eqnarray}
where $Q_\pm(t)=\int_t^{+\infty} P_\pm(t')\,dt'$ is the probability
that a $\pm$ interval is larger than $t$. For even $N=2n$ $(n\geq
1)$, one obtains a similar expression
\begin{eqnarray}
&P_<(2n,t)={\tau_-^{-1}}\int_0^t dt_1\,Q_-(t_1){\times}
\nonumber\\
&\int_{t_1}^t dt_2\,P_+(t_2-t_1) \int_{t_2}^t
dt_3\,P_-(t_3-t_2)\cdots \nonumber\\
&\int_{t_{2n-2}}^t dt_{2n-1}\,P_-(t_{2n-1}-t_{2n-2}){\times}
\nonumber\\
&\int_{t_{2n-1}}^t
dt_{2n}\,P_+(t_{2n}-t_{2n-1})Q_-(t-t_{2n}).\label{conv2}
\end{eqnarray}
For any function of time $F(t)$ introduced in this letter, one
defines its Laplace transform $\hat F(s)=\int_0^{+\infty}F(t)\,{\rm
e}^{-st}\,dt$. The convolution products in
Eqs.~(\ref{conv1},\ref{conv2}) take a much simpler form in the
Laplace variable $s$
\begin{eqnarray}
\hat P_<(2n-1,s)&=&{\tau_-^{-1}}\hat Q_+\hat Q_- [\hat P_+\hat
P_-]^{n-1},\label{podd}\\
\hat P_<(2n,s)&=&{\tau_-^{-1}}\hat Q_-^2P_+ [\hat P_+\hat
P_-]^{n-1},\label{peven}
\end{eqnarray}
where $\hat Q_\pm(s)=\frac{1-\hat P_\pm(s)}{s}$. One can now
express the conservation of probability, $
P_<(t)+\sum_{N=1}^{+\infty} P_<(N,t)=1, $ which leads to
\begin{equation}
\hat P_<(s)=\frac{1}{s}-\frac{1-\hat P_-(s)}{\tau_- \,s^2}.
\label{pinf}
\end{equation}
In fact, Eq.~(\ref{pinf}) is an {\it exact} relation, which reads
\begin{equation}
P_<(t)=\tau_-^{-1}\int_t^{+\infty}(t'-t)P_-(t')\,dt',
\label{pinfnew}
\end{equation}
in the time variable. Indeed, if $X(t)$ has not crossed the level
$M$ up to time $t$, it belongs to a $-$ interval of duration
$t'>t$, starting at an initial position uniformly distributed between 0 and
$t'-t$. Note that $\hat P_>(s)$ and $\hat P_>(N,s)$ are given by
similar expressions as Eqs.~(\ref{podd},\ref{peven},\ref{pinf}) by
exchanging the indices $-$ and $+$.

Now, $\hat P_\pm(s)$ can be calculated by expressing the {\it
known} quantities $\hat A(s)$ and $\hat N_<(s)$ as a function of
$\hat P_\pm(s)$:
\begin{eqnarray}
\hat N_<(s)=\frac{(1+\hat P_+)(1-\hat P_-)}{\tau_- \,s^2(1-\hat
P_+\hat P_-)},\label{eq0}\\
\hat A(s)=G(M)\left[\frac{1}{s}-\frac{1-\hat P_+}{1+\hat P_+}\,
N_<(s) \right].\label{eq1}
\end{eqnarray}
Using $\hat P_\pm'(0)=-\tau_\pm$ and Eq.~(\ref{t2}), one obtains
the following estimates, valid for small $s$,
\begin{equation}
\hat N_<(s)\sim\frac{1}{\tau \,s^2},\quad\hat
A(s)\sim\frac{G^2(M)}{s}.\label{est}
\end{equation}
The first expression in Eq.~(\ref{est}) is equivalent to
Eq.~(\ref{as}), whereas the second relation expresses that for large
$t$, $A(t)\sim G^2(M)$. For large $s$, $\hat N_<(s)\sim [2G(M)\tau
s^2]^{-1}$, which corresponds to the small time behavior
\begin{equation}
N_<(t)\sim\frac{t}{2G(M)\tau},\quad
N_>(t)\sim\frac{t}{2(1-G(M))\tau}.\label{smalln}
\end{equation}
For $G(M)\ne \frac{1}{2}$ ({\it i.e.} $M\ne 0$, when $g(X)$ is
symmetric), Eq.~(\ref{smalln}) differs from the large time
asymptotics given by Eq.~(\ref{as}). However, the sum rule of
Eq.~(\ref{sumrule}) is preserved by the small time estimates of
Eq.~(\ref{smalln}). Finally, writing
\begin{equation}
\hat F(s)=\frac{G(M)-s\,A(s)}{G(M)\,sN_<(s)},\label{eq2}
\end{equation}
and using Eqs.~(\ref{eq0},\ref{eq1}), the interval distributions
read
\begin{eqnarray}
\hat P_+(s)&=&\frac{1-\hat F(s)}{1+ \hat F(s)}, \label{pp} \\
\hat P_-(s)&=&\frac{2-\tau_- \,s^2N_<(s)(1+\hat F(s))}
{2-\tau_-\,s^2N_<(s)(1-\hat F(s))}.\label{pm}
\end{eqnarray}
Inserting these expressions of $P_\pm$ in Eq.~(\ref{pinf}), one
obtains our final result for $P_<$ (and $P_>$), from the sole
knowledge of $A(t)$ and $N_<(t)$ (or $N_>(t)$).

The persistence exponent $\theta$ is defined as the asymptotic decay
rate of $P_<(t)\sim {\rm e}^{-\theta t}$. The term ``exponent''
arises from the fact that in many physical systems
\cite{per1,per2,iia,potts,global,krug,breath,lq,soap,diff1d,surf},
the process $X$ of interest is stationary in the variable $t=\ln T$,
where $T$ is the actual physical time. Thus the persistence decays
as a power law $P_<(T)\sim T^{-\theta}$, as a function of the real
time $T$ (see below for practical examples). Within our approach,
$-\theta$ is the first pole of $\hat P_<(s)$ (or equivalently of
$\hat P_-(s)$) on the negative real axis \cite{iia}. Using
Eq.~(\ref{pm}), one finds that $\theta$ satisfies the implicit
equation
\begin{equation}
\theta G(M)[1+\theta N_<(-\theta)]+ \theta^2
A(-\theta)=\tau^{-1}.\label{thetafina}
\end{equation}
When $M$ is large, $\theta$ goes to zero, and Eq.~(\ref{thetafina})
leads to
\begin{equation}
\theta=\tau_-^{-1}=(2\tau G(M))^{-1}.\label{thetaana}
\end{equation}
For a Gaussian process, the same expression was obtained from a
heuristic argument in \cite{merca}. In this limit of large $M$, the
interval distributions are found to become Poissonian. We conjecture
that the present approach becomes exact for large $M$, the $-$
intervals being so large that the $+$ intervals are indeed
uncorrelated.

Let us move on to the case where $X(t)$ is a stationary {\it
Gaussian} process. The properties of $X(t)$ (and hence $A(t)$ and
$N_<(t)$) are completely determined by the sole knowledge of its
two-time correlator $f(t)=\langle X(0)X(t)\rangle$. For a general
process, this connection is only approximate and can only be made by
means of the IIA \cite{iianewpaper}. For convenience, we set
$\langle X^2(t)\rangle=f(0)=1$. The process is smooth if $f$ is
twice differentiable. We also assume that for large time $t$, the
correlator $f(t)$ decays fast enough so that
$\int_0^{\infty}f(t)\,dt$ is finite \cite{BL,iia}. For $t>0$, the
position-velocity correlator is $\langle X(0)X'(t)\rangle=f'(t)$,
which vanishes for $t=0$, since $f(t)$ is an even function, twice
differentiable at $t=0$. The velocity-velocity correlation function
is $\langle X'(0)X'(t)\rangle=-f''(t)$. The mean time interval
between crossings $\tau$ is computed using Eq.~(\ref{taugen})
\begin{equation}
\tau^{-1}=\frac{\sqrt{-f''(0)}}{\pi}{\rm e}^{-\frac{M^2}{2}}.
\end{equation}
For a Gaussian process, $A(t)$ has been derived in \cite{potts}
\begin{equation}
A(t)=\int_{-\infty}^M g(x)\,G\left(\frac{M-x
f(t)}{\sqrt{1-f^2(t)}}\right)\,dx.
\end{equation}
For large time, so that $f(t)$ is small, one finds
\begin{equation}
A(t)=G^2(M)+\frac{f(t)}{2\pi}{\rm e}^{-M^2}+O(f^2(t)).
\end{equation}
Finally, $N_<(t)$ can be calculated after introducing the
correlation matrix of the Gaussian vector $(X(t),X(0),X'(t))$,
which reads
\begin{equation}
{\cal C}(t)=\left(\begin{array}{ccc}
   1 & f(t) & 0 \\
   f(t) & 1 & f'(t) \\
   0 & f'(t) & -f''(0)
\end{array}\right).
\end{equation}
One finds
\begin{equation}
N_<(t)=G^{-1}(M)\int_{0}^{t}\langle |X'(t')| \rangle_{<}\,dt',
\end{equation}
where $\langle |X'(t)| \rangle_{<}$ is the average of the velocity
modulus, knowing that $X(t)=M$, and averaged over $X(0)<M$:
\begin{equation}
\langle |X'(t)| \rangle_{<}=\int_{-\infty}^{M}dx_0
\int_{-\infty}^{+\infty}dv \,\frac{|v|{\rm e}^{-\frac{1}{2}{\bf
U}^\dag{\cal C}^{-1}{\bf U}}}{(2\pi)^{3/2}\sqrt{\det C}},
\end{equation}
where ${\bf U}=(M,x_0,v)$. In practice, this integral has to be
computed numerically, but can be reduced to a cumbersome
one-dimensional integral over $x_0$, involving $g$ and $G$. For
large time or small $f(t)$, one obtains
\begin{equation}
\langle |X'(t)| \rangle_{<}
-\frac{1}{\tau}=-\frac{Mg(M)}{G(M)\tau}f(t)+O(f^2(t)).\label{largen}
\end{equation}
Note that for numerical purposes, the Laplace transform of
$N_<(t)$ can be efficiently written as
\begin{equation}
\hat N_<(s)=\frac{1}{s}\int_0^{+\infty}\left(\langle |X'(t)|
\rangle_{<}- \frac{1}{\tau}\right){\rm e}^{-st}\,dt+\frac{1}{\tau
\,s^2}.
\end{equation}
The various analytical asymptotic forms obtained above can be useful
to complement the partial knowledge of $A(t)$ and $N_<(t)$ from a
partial experimental or numerical sampling of the process $X(t)$
\cite{iianewpaper}.

As an application, Table~\ref{tab1} reports the theoretical and
numerical values  (the latter obtained by direct simulation of the
temporal process) of the persistence exponent $\theta$ for
$M=0,\,1,\,2,\,3$ and for the two Gaussian processes $X_1$ and
$X_2$ associated to the correlators
\begin{eqnarray}
f_1(t)&=&\frac{1}{\cosh\left(\frac{t}{2}\right)},\\
f_2(t)&=&\frac{1}{2}\left(3{\rm e}^{-\frac{|t|}{2}}-{\rm
e}^{-\frac{3|t|}{2}}\right).
\end{eqnarray}
$X_1$ and $X_2$ are two examples of non-Markovian Gaussian processes
arising from physical systems. Indeed, up to a multiplicative term,
$X_1(t)$ can be shown to be equal to $\rho({\bf x},T)$ (at some
arbitrary position ${\bf x}$), where the density field $\rho({\bf
x},T)$ evolves according to the two-dimensional diffusion or heat
equation \cite{iia}, $\frac{\partial\rho}{\partial T}=\nabla^2\rho,$
starting from an arbitrary (although not too strongly correlated)
initial condition. Here, the actual time $T$ is again related to our
stationary time $t$ by the relation $t=\ln T$. Note that the out of
equilibrium dynamics of a two-dimensional Ising model after a quench
at a temperature $T_0<T_c$ (where $T_c$ is the ferromagnetic
critical temperature) can be approximately mapped to this problem
\cite{iia,potts}, by assimilating the spin $S({\bf x},T)=\pm 1$ to
$S({\bf x},T)={\rm sign}[\rho({\bf x},T)]$. We find that the
theoretical values are within the numerical error bars, except maybe
for $M=0$. Overall, the accuracy is better than 1\%. In addition,
the asymptotic result of Eq.~(\ref{thetaana}) already leads to fair
estimates for $M=2$ and $M=3$ ($\theta_1(M=2)=1.102\cdot 10^{-2}$
and $\theta_1(M=3)= 8.852\cdot 10^{-4}$). Our theoretical and
numerical results are also consistent with the numerical bounds
computed in \cite{merca}, for $M=1$ and $M=2$
($0.0586<\theta_1(M=1)<0.0684$ and $0.0106<\theta_1(M=2)<0.0119$).
\begin{table}
\begin{tabular}{ccccc}
$M$&$\theta_{1}^{th}$ & $\theta_{1}^{sim}$ &$\theta_{2}^{th}$ &
$\theta_{2}^{sim}$\\ \hline 0 & $0.1862$ & 0.188(1) & 0.2647 & 1/4 \\
1 & $5.914\cdot 10^{-2}$ & $5.91(1)\cdot 10^{-2}$ & $8.625\cdot
10^{-2}$ & $8.31(4)\cdot 10^{-2}$ \\ 2 & $1.084\cdot 10^{-2}$ &
$1.09(1)\cdot 10^{-2}$ & $1.665\cdot 10^{-2}$ & $1.61(2)\cdot
10^{-2}$\\ 3 & $8.769\cdot 10^{-4}$ & $8.77(2)\cdot 10^{-4}$ &
$1.420\cdot 10^{-3}$ & $1.42(1)\cdot 10^{-3}$ \\
\end{tabular}
\caption{\label{tab1}Exponents $\theta$ from theory
($\theta^{th}$) and simulations ($\theta^{sim}$), for different
values of $M$, calculated for the processes $X_1$ and $X_2$
introduced in the text. For $M=0$, the three first results were
first reported in \cite{iia}, while the exact value
$\theta_{2}(M=0)=\frac{1}{4}$ was obtained in \cite{theta2}.}
\end{table}

As for $X_2$, it is associated with the random acceleration process
\cite{theta2}, $\frac{d^2X}{dT^2}=\eta(T)$, where $\eta(T)$ is a
$\delta$-correlated white noise, in the variable $T={\rm e}^t$.
Contrary to $X_1$, the process $X_2$ is not infinitely
differentiable ($f_2'''(t)$ is not defined at $t=0$), although it is
just smooth enough for the present approach to be applicable. Hence,
it is not surprising that the theoretical results are not as good as
for the process $X_1$. However, the theoretical estimates are
clearly becoming more accurate as $M$ increases, and presumably
exact for large $M$.

In conclusion, the present work develops a powerful approximation
leading to explicit expressions for the Laplace transform of the
probability to remain above or below a certain level $M$ (and hence
the distribution of the minimum or maximum of the process). This
approach also gives the distribution of time intervals during which
the process remains above or below $M$, and leads to the
determination of the persistence exponent.

\acknowledgments I am very grateful to Satya Majumdar and Partha
Mitra for fruitful discussions.

\end{document}